\documentstyle[11pt,aasms4,rotating]{article} %\documentstyle[11pt,aaspp4]{article}

\def\simlt{\lower.5ex\hbox{$\; \buildrel < \over \sim \;$}}
\def\simgt{\lower.5ex\hbox{$\; \buildrel > \over \sim \;$}}

\def\gcm3{{\rm\,g\,cm^{-3}}}
\def\ncm3{{\rm\,cm^{-3}}}

\def\>{$>$}
\def\<{$<$}

\lefthead{Liu et al.}
\righthead{}

\begin{document}
\centerline{Submitted to the Editor of the Astrophysical Journal Letters}
\vskip 0.5in
\title{\bf Accretion Processes in the Nucleus of M31}

\author{Siming Liu\altaffilmark{1} and Fulvio Melia$^{1,2,3}$}

\affil{$^1$Physics Department, The University of Arizona, Tucson, AZ 85721}
\affil{$^2$Steward Observatory, The University of Arizona, Tucson, AZ 85721}

% Notice that each of these authors has alternate affiliations, which
% are identified by the \altaffilmark after each name.  The actual alternate
% affiliation information is typeset in footnotes at the bottom of the
% first page, and the text itself is specified in \altaffiltext commands.
% There is a separate \altaffiltext for each alternate affiliation
% indicated above.

\altaffiltext{3}{Sir Thomas Lyle Fellow and Miegunyah Fellow.}

% The abstract environment prints out the receipt and acceptance dates
% if they are relevant for the journal style.  For the aasms style, they
% will print out as horizontal rules for the editorial staff to type
% on, so long as the author does not include \received and \accepted
% commands.  This should not be done, since \received and \accepted dates
% are not known to the author.

\begin{abstract}
The hypothesized supermassive black hole in the nucleus of M31 (which
we shall hereafter call M31*) has many features in common with Sgr A* 
at the Galactic Center, yet they differ in several significant and important 
ways. Though M31* is probably ten times heavier, its radio luminosity
at 3.6 cm is only one third that of Sgr A*.  At the same time, M31* is
apparently thousands of times more luminous in X-rays
than its Galactic Center counterpart.
Thus, a comparative study of these objects can be valuable in helping
us to understand the underlying physical basis for their activity.
We show here that the accretion model being developed for Sgr A*
comprises two branches of solutions, distinguished by the relative importance
of cooling compared to compressional heating at the radius $r_C$ where the
ambient gas is captured by the black hole.  For typical conditions in the
ISM, the initial temperature ($T[r_C]\sim 10^6-10^7$ K) sits on the
unstable branch of the cooling function.  Depending on the actual
value of $T(r_C)$ and the accretion rate, the plasma settles either onto
a hot branch (attaining a temperature as high as $10^{10} K$ or so at
small radii) or a cold branch, in which $T$ drops to $\sim 10^4$ K.
Sgr A* is presumably a `hot' black hole. We show here that the VLA,
UV and {\it Chandra} observations of M31* reveal it to be a member of
the `cold' black hole family. We discuss several predicted 
features in the spectrum of M31*
that may be testable by future multi-wavelength observations, including
the presence of a prominent UV spike (from hydrogen line emission) that
would be absent on the hot branch.
\end{abstract}

% The different journals have different requirements for keywords.  The
% keywords.apj file, found on aas.org in the pubs/aastex-misc directory, 
% contains a list of keywords used with the ApJ and Letters.  These are 
% usually assigned by the editor, but authors may include them in their 
% manuscripts if they wish. 


% That's it for the front matter.  On to the main body of the paper.
% We'll only put in tutorial remarks at the beginning of each section
% so you can see entire sections together.

% In the first two sections, you should notice the use of the LaTeX \cite
% command to identify citations.  The citations are tied to the
% reference list via symbolic KEYs.  We have chosen the first three
% characters of the first author's name plus the last two numeral of the
% year of publication.  The corresponding reference has a \bibitem
% command in the reference list below.
%
% Please see the AASTeX manual for a more complete discussion on how to make
% \cite-\bibitem work for you.   

\section{Introduction}

The discovery of an unresolved radio point source (labeled as M31*) in
the nucleus of M31 (Crane, Dickel \& Cowan 1992) led to an early suggestion
(Melia 1992b) that its nature may be similar to that of the supermassive black hole
candidate, Sgr A*, at the center of our own Galaxy.  The distance to M31
renders its nuclear environment difficult to observe with anything approaching the
spatial resolution now available for the Galactic Center, but certain 
cues suggest that in both cases we may be dealing with a massive pointlike
object accreting gas from the interstellar medium, possibly from stellar winds.
However, there do exist several notable differences between the two systems, 
making a comparative analysis of their emitting regions highly desirable. 

The {\it Hubble Space Telescope} (HST) resolved the nucleus of M31 into two 
components (P1 and P2) separated by $0\farcs5\approx
1.9$ pc (Lauer et al. 1993). The rotational velocity measurements of the 
stars near the nucleus (Kormendy \& Bender 1999) suggest that this
structure is a torus of stars orbiting a central dark, object of mass 
$3.0\times 10^7\, M_\odot$ in a slightly eccentric trajectory (Tremaine 1995). 
High resolution UV images of the nucleus show that there is a 
group of UV-bright stars between P1 and P2 (Brown et al. 1998), making it very
difficult to study the compact object at this frequency. A pre-Costar HST
far-UV observation in 1991 (King, Stanford \& Crane 1995) could not
resolve this cluster from P2, but it did show that the region near P2 has
a UV upturn that is much stronger than P1 and the
surrounding bulge light.  These authors argue that the FUV flux level 
($\sim 3\mu$Jy) from P2 is not from stars, but is rather 
a high-frequency extension of the radio source M31*.  This view is supported
by the arguments of Kormendy and Bender (1999) that the central super-massive
black hole probably lies at the center of this UV-bright cluster.

In the radio, there now exist measurements of M31*'s flux at two epochs 
(Crane et al. 1992; Crane et al. 1993), showing that, at least at $3.6$ 
cm, its luminosity is variable on a time scale of a year or less.  But even 
though M31* is apparently ten times more massive than the black hole at the 
Galactic Center, its power at $3.6$ cm is only one third that of Sgr A* (assuming 
a distance to M31 of $784$ kpc; Stanek \& Garnavich 1998).  This contrasts
with the situation at higher energy, where M31* is thousands of 
times more luminous than the latter in X-rays (Garcia et al. 2000). The recent 
{\it Chandra} observation also shows that M31* has a soft X-ray 
spectrum, while the X-ray spectrum of Sgr A* may be harder (Baganoff et al. 
2001; Melia, Liu, \& Coker 2000b).  

The recent modeling of Sgr A* (Melia, Liu \& Coker 2000a, 2000b), taking 
into account the latest high precision mm, sub-mm and {\it Chandra} measurements, 
suggests that (at least in this source) the rate of accretion ($\dot M$) toward 
small radii ($< 30\,r_S$, where $r_S$ is the Schwarzschild radius $\equiv 
2GM/c^2$) must be significantly smaller than the rate of wind capture at the 
Bondi-Hoyle radius $r_C\equiv 2GM/v_\infty^2$, where $v_\infty$ is the velocity 
of the ambient gas flowing past the central object. Hydrodynamic simulations 
of the environment surrounding Sgr A* (Coker \& Melia 1997) indicate that the infalling gas 
circularizes when it approaches within $5-25\;r_S$ of the black hole.
The sub-mm ``excess'' emission seen in the spectrum of this source appears to
be associated with radiation produced within the inner Keplerian region.
This view is supported by the linear polarization observed in Sgr A* at mm and 
sub-mm wavelengths (Aitken et al. 2000), which is difficult to reconcile with 
other geometries of the emitting region.   Both the degree of polarization
and the rotation (by about $90^\circ$) in the position angle appear to
be consistent with a transition of the emitting region within the Keplerian
flow from optically thick emission (at mm and longer wavelengths) to optically 
thin emission in the sub-mm range (Melia 1992a; 1994).  The overall
spectrum from this region includes a high-energy component due to 
bremsstrahlung and inverse Compton scattering processes, which may account
for the {\it Chandra} X-ray source coincident with Sgr A* if it turns out 
to be the actual counterpart to this object (Baganoff et al. 2001).

The situation in the nucleus of M31 may be quite different for several 
reasons.  First, $v_\infty$ around Sgr A* may be anomalously high due to the
presence of strong wind-producing young stars (e.g., Najarro et al. 1997).  
In M31, on the other hand,
very little is known about the nuclear gas motion, although on a larger scale, 
Rubin \& Ford (1971) find that the plasma within the inner $400$ pc region 
shows a complex velocity field superposed on the rapid rotation, with evidence
for expansion at velocities up to $\sim 100$ km s$^{-1}$.  Second, the central
stellar distribution, as noted above, is clearly anisotropic, with P2 lying
a mere $0\farcs1$ ($\approx 1.3\times 10^5\,r_S$) away from 
the central black hole (Kormendy \& Bender 1999).  In M31, a
lower $v_\infty$, together with the disruption to the gas flow caused by these
stars, may result in a greater $\dot M$ with a smaller specific angular
momentum.

It is well known that the gas temperature $T$ near $r_C$ depends
on the kinetic energy flux brought into this region by the pre-shocked gas 
(see Coker \& Melia 2000).
An important consequence of a weaker flow is a value of
$T$ as low as $10^6$ K at large radii, compared with $\sim 10^7$ K 
in the Galactic Center. A thin plasma under these conditions sits on the unstable
branch of the cooling function (Gehrels \& Williams 1993). As it continues
to fall toward the accretor, it can either heat up to $>10^{10}$ K, 
at which point the cooling rate balances heating, or it cools down to 
$\sim 10^4$ K, where the energy loss rate drops precipitously. In general, a lower 
initial temperature and a larger particle number density force the plasma into
the lower $T$ stable branch while the higher $T$ branch is favored by the gas
starting its descent with a high $T$ and low density. 

In this {\it Letter}, we explore this dichotomy in accretion profiles, and
suggest that Sgr A* represents the hot branch, while M31* is 
a member of the cold branch family of accreting black holes. In both cases,
the initial temperature near $r_C$ falls within the unstable region,
but M31* is accreting at a higher rate than Sgr A*.  If indeed 
the plasma in M31* is stabilized at $\sim1.5\times 10^4$ K, its FUV spectrum 
should be characterized by strong H line emission, while the soft X-rays 
are produced by recombination in the hot outer region.  This is in contrast with 
Sgr A*, where the X-rays are produced by the inner hot gas.  In addition, we 
inferred a mass loss for Sgr A* to account for the variable $\dot M$ toward 
small radii.  This may not occur in M31* where the inflowing gas 
is much colder and hence more tightly bound (see Fig. 1).  We emphasize 
that our analysis and conclusions regarding M31* are based on the assumption 
that both the $3.6$ cm and {\it Chandra} measurements represent the actual 
source strength, rather than upper limits. One of our goals is to make available 
several testable predictions for the next generation of coordinated observations.

\section{Accretion from the Ambient Medium and Calculation of the Spectrum}
The currently available spectral measurements for the nuclear region in M31 are
listed in the Table. The column labeled ``Integration Area'' shows the region over 
which the intensity was integrated in order to arrive at the flux measurement 
quoted in column 3.  The angular resolution for the 3.6 cm observations was 
$0\farcs42\times 0\farcs31$ and $0\farcs29\times 0\farcs26$, respectively, which is
sufficient to attribute the observed flux to a single point source. The FOC (on 
{\it HST}) observations of M31* are hampered by contamination from the blue stars around
the center (Brown et al. 1998), so these should be considered as upper limits.
However, the $0.175\mu$m point represents the flux from this region once the 
contribution from these stars has been subtracted out (King et al. 1995).
With a spatial resolution of $0\farcs8$, {\it Chandra} was able to resolve the
nucleus of M31 into five point sources (Garcia et al. 2000). These authors 
infer that the X-ray counterpart to M31* is strongly variable, when the recent data 
are compared to earlier {\it Einstein} observations of the TF 56 source 
(Trinchieri \& Fabbiano 1991).

A direct application of the Sgr A* accretion model (e.g., Coker \& Melia 2000;
Melia, Liu \& Coker 2000b) to M31* immediately runs into problems because of 
the large disparity between the radio to X-ray flux ratios of these two sources.
The spectrum of M31* simply does not support the idea that the inflowing gas
is hot ($\sim 10^{10}$ K or so) at small radii.  We have therefore explored
the characteristics of the accretion model in the high-$\dot M$ context to gauge
the importance of rapid cooling in this environment.  

The relativistic black hole accretion model was pioneered by Shapiro (1973),
and later developed by several workers, the more recent of whom applied
this picture to Sgr A* (e.g., Melia 1992a, 1994). Our calculation of the accretion 
profile for M31* follows the prescription of Coker \& Melia
(2000), though with several important differences.  First, we will restrict our
attention to the cold branch solutions for which the impact of a magnetic field $B$
is negligible.  This assumes that $B$ is sub-equipartition compared to the
thermal energy density, following the arguments of Kowalenko \& Melia (1999),
and the application to Sgr A* in the outer region (Coker \& Melia 2000). 
(Here the magnetic pressure never exceeds $\sim 2\times 10^{-2}$ times
the thermal pressure.) Thus, it is not important here to worry about an empirical 
fit for $B$, as we did for Sgr A*.  
Second, we are in a domain where cooling dominates the energy equation,
especially at large radii where the conditions warrant a careful treatment of
line cooling (Gehrels \& Williams 1993).  Thus, whereas this effect could be
ignored in the case of Sgr A*, we must include it here, and for this we adopt 
the cooling function used by these authors, corresponding to a thin gas with
cosmic abundances. 

\small{
\begin{quote}
\begin{tabular}{ccccccc}
\hline
\hline
$\lambda$       & $\nu$  & $F_\nu$ &Integration & Telescope &
Date & Notes\\
or Energy band    &  (Hz)  &   (Jy)  &Area      & or Instrument&      &
\\
\hline
$3.6$cm  & $8.4\times 10^9$ & $(28\pm 5)\times 10^{-6}$ &
$0\farcs42\times 0\farcs31$ & VLA & June 1990 & 1 \\
$3.6$cm  & $8.4\times 10^9$ & $(39\pm 5)\times 10^{-6}$ &
$0\farcs29\times 0\farcs26$ & VLA & Nov 1992  & 2 \\ 
$100\mu$m& $3.0\times 10^{12}$ & $12.0$ & $2\farcm0\times
2\farcm0$ & IRAS & 1983 & 3,4\\
$60\mu$m & $5.0\times 10^{12}$ & $7.1$  & $2\farcm0\times
2\farcm0$ & IRAS & 1983 & 3,4\\
$25\mu$m & $1.2\times 10^{13}$ & $0.91$ & $2\farcm0\times
2\farcm0$ & IRAS & 1983 & 3,4\\
$12\mu$m & $2.5\times 10^{13}$ & $1.83$ & $2\farcm0\times
2\farcm0$ & IRAS & 1983 & 3,4\\
$0.75\mu$m & $4.0\times 10^{14}$ & 0.746 & $190^\prime\times
60^\prime$ & 0.9m(KPNO) & Sep 1991 & 5 \\
$0.61\mu$m & $5.0\times 10^{14}$ & 0.417 & $190^\prime\times
60^\prime$ & 0.9m(KPNO) & Sep 1991 & 5 \\
$0.53\mu$m & $5.6\times 10^{14}$ & 0.278 & $190^\prime\times
60^\prime$ & 0.9m(KPNO) & Sep 1991 & 5 \\
$0.46\mu$m & $6.5\times 10^{14}$ & 0.177 & $190^\prime\times
60^\prime$ & 0.9m(KPNO) & Sep 1991 & 5 \\
$0.40\mu$m & $7.5\times 10^{14}$ & 0.075 & $190^\prime\times
60^\prime$ & 0.9m(KPNO) & Sep 1991 & 5 \\
$0.35\mu$m & $8.6\times 10^{14}$ & 0.024 & $190^\prime\times
60^\prime$ & 0.9m(KPNO) & Sep 1991 & 5 \\
$0.280\mu$m& $1.1\times 10^{15}$ & $4.85\times 10^{-4}$ &
$2\farcs8 \times 2\farcs8$ & FOC on HST & Feb 1994 & 6 \\
$0.199\mu$m &$1.5\times 10^{15}$ & $2.47\times 10^{-4}$ &
$2\farcs8 \times 2\farcs8$ & FOC on HST & Feb 1994 & 6 \\
$0.175\mu$m &$1.7\times 10^{15}$ & $6.7\times 10^{-6}$ &
$0\farcs5 \times 0\farcs5$ & FOC on HST & 1991 & 7 \\
0.3-1.5keV  &$(0.7-3.6)\times 10^{17}$ & $1.8\times 10^{-7}$ &
$2\farcs0 \times 2\farcs0$ & Chandra & Oct 1999 & 8 \\
1.5-7.0keV  &$(0.4-1.7)\times 10^{18}$ & $7.8\times 10^{-10}$ &
$2\farcs0 \times 2\farcs0$ & Chandra & Oct 1999 & 8\\
\hline
\end{tabular} 

\hspace{0.5cm}Notes: (1) Crane, Dickel \& Cowan. 1992. (2) Crane et
al. 1993. (3) Soifer et
al. 1986. (4) Neugebauer et al. 1984. (5) Mcquade, Calzetti \& Kinney
1995. (6) Brown  et al. 1998. (7) King, Stanford \& Crane. 1995. (8) Garcia et
al. 2000.
\end{quote}
}

\normalsize

Without a specific determination of the gas conditions within a 
parsec or so of M31*, we are compelled to treat the gas kinematics as an 
unknown. We therefore adopt a fiducial wind velocity $v_\infty$ of $500$ 
km s$^{-1}$.  In the cold branch solutions, a lower value of $v_\infty$ 
simply cools the gas faster because $\dot M$ is higher.  The corresponding 
Bondi-Hoyle capture radius for M31* is then $r_C=3.6\times 10^5 r_S$ 
($\approx 0\farcs28$), which by the way, is larger than the distance from
M31* to P2 (see above). With $M$ and $v_\infty$ known, the accretion rate 
then depends on the gas density, which unfortunately is also poorly 
constrained.  However, Ciardullo et al. (1988) argue that the ionized-gas 
density gradient in the nucleus of M31 is 
dramatic, dropping from $n_e\sim 10^4$ electrons cm$^{-3}$ at a radius of 
$\sim 7^{\prime\prime}$ to $10^2$ electrons cm$^{-3}$ at $\sim 1^{\prime}$.
Thus, simply putting $\dot M\sim \pi r_C^2\,m_pn_e\,v_\infty$, where $m_p$ is
the proton mass, and $n_e=10^4$ cm$^{-3}$, we see that $\dot M$ may be as high as 
$10^{25}$ g s$^{-1}$.

The temperature at the outer boundary ($\sim r_C$) is estimated by assuming that  
the kinetic energy in the gas is thermalized by shocks, for which 
$6 R_g T(r_C)\sim (3v_\infty/4)^2$, where $R_g$ is the gas constant and we have 
assumed a mean molecular weight per particle of $1/2$.  The principal radiating 
mechanisms are radiative recombination, and thermal bremsstrahlung emission. 
Electron-positron pair production, cyclo/synchrotron 
radiation and Comptonization are all negligible due to the low temperature and
sub-equipartition magnetic field in the cold branch solutions. 
For simplicity, we here consider only H recombination to calculate the line
emission.  A more sophisticated treatment that includes the contribution from all
the ions will be discussed elsewhere.  For this, the ionization fraction may
be calculated according to the prescription given in Rossi et al. (1997).
At the prevalent temperature ($\sim 1.4\times 10^4$ K) in the cold region,
ignoring the other line emissivities amounts to an error of at most 
$\sim 15\%$ (Brown \& Mathews 1970).

\section{Results and Conclusions}
The principal distinguishing feature between the hot and cold branch solutions
is illustrated in Figure 1, which shows the temperature profile of the accreting
gas as a function of radius.  The emphasis here is the dependence of this
profile on the initial temperature, so all the other physical conditions are
identical for the three cases included on this plot, and are based on the best
fit model for the spectrum of M31* (dark, solid curve).  In all cases, the outer
boundary is taken to be $r_0= 10^5\,r_S$ ($\approx 0\farcs08$), 
with an accretion rate
$\dot M=1.5\times 10^{24}$ g s$^{-1}$.  The ratio of thermal to gravitational
energy density at $r_0$, which characterizes the initial temperature, is
$0.19$, $0.17$ and $0.15$, respectively, for the thin, solid curve (the sole
hot branch solution), the thick, solid curve (the best fit model), and 
the dotted curve.  With an appreciably lower $\dot M$,
Sgr A* lies comfortably within the category of hot branch solutions since its
initial temperature appears to be larger than that of M31* (at $\sim 3\times
10^6$ K).  For the latter, however, the gas cools
quickly as it collapses towards smaller radii, and has reached the
cold, stable branch of the cooling curve (at $\sim 10^4$ K) by the time it 
crosses $\sim 10^3\, r_S$. Thereafter, the heating due to compression 
never quite catches up and the gas continues
to radiate efficiently as it plummets toward the event horizon.

The spectrum produced by this plasma is shown in Figure 2 (dark, solid curve),
in which we have also indicated the contribution (mostly due to bremsstrahlung
emission) from the hot ($T> 10^6$ K), outer region (thin, solid curve), and 
the cooler ($T<10^6$ K), inner zone
(dotted curve). The transition radius between these two domains is approximately
$10^3\,r_S$ (see Fig. 1).  The most constraining data on this plot are the VLA 
measurements (Crane et al. 1992; Crane et al. 1993), the FUV point inferred from 
the subtraction of starlight (King et al. 1995), and the more recent {\it Chandra} 
results (shown as a butterfly below $1.5$ keV and as an upper limit above this 
energy; Garcia et al. 2000).  We see that the X-rays are produced in the outer,
hot region, whereas the FUV spike arises from hydrogen line emission, predominantly 
in the cooler, inner zone. The radio portion of the spectrum is produced throughout
the accretion volume.

We would thus expect M31*'s X-ray flux to vary on a time scale
$\tau\sim 10^3\,r_S/v_{ff}\le 4$ months, where the free-fall velocity
$v_{ff}$ at $10^3\,r_S$ is approximately $c/30$.  Its radio flux, which
is produced throughout the volume of captured gas, should also vary on a
time scale no larger than this, though only a portion of this variability
is expected to be correlated with the X-rays.  Since some of the radio
emission occurs at smaller radii, we expect that relative to the X-rays,
the 3.6 cm emission ought to display variability with a broader range
of time scales.

Figure 2 also highlights one of the main features that distinguish
the spectra of the hot and cold solutions---the FUV spike due to line 
emission, which is absent
in the former, but is very prominent in the latter.  This may be the origin of
the peculiar UV upturn near P2 noted by King et al. (1995) in their analysis of
the non-stellar contribution to the UV flux.  The strength of this spectral
component suggests that additional UV observations of the region near P2 are
called for.  The variability in the UV flux appears to be correlated
with both the flux at 3.6 cm, and the X-rays, though the tightest 
correlation is expected to occur between the latter two.

In recent years, several other models have been invoked to account for the
emissivity of supermassive black holes in the cores of nearby galaxies, including
Sgr A*.  ADAF (Narayan, Yi, \& Mahadevan 1995) and ADIOS (Blandford \&
Begelman 1999) disk models differ from the model we have described here in 
several ways, including (1) the accreting gas has considerably more angular 
momentum than we have invoked, and (2) the gas separates into a two-temperature 
plasma, in which the protons become very hot (analogous to our hot-branch 
solution). However, the VLA point appears to rule out this possibility in the 
case of M31*, for the simple reason that an optically thick emitter at this 
temperature, with a scale size corresponding to $M=3\times 10^7\;M_\odot$, 
would simply produce too much flux at 3.6 cm.  In addition, a 
hot, optically-thin emitter does not produce the correct X-ray spectrum.  
These constraints also seem to argue against a jet model (Falcke \& Markoff 
2000), since the emitting particles in this picture would have physical 
characteristics similar to those of our hot-branch solution. The combination
of radio and X-ray measurements constitutes a powerful diagnostic.  Future 
co-ordinated multi-wavelength observations of this intriguing supermassive 
black hole candidate would be invaluable.

We are indebted to Fred Baganoff, Mike Garcia, Ivan King, and Mark Morris
for very valuable discussions and clarifications. This work was supported by a 
Sir Thomas Lyle Fellowship and a Miegunyah Fellowship in 
Melbourne, and by NASA grants NAG5-8239 and NAG5-9205.

{}

% And finally, we must deal with the figures.  There are three figures
% associated with this manuscript; two figures are Encapsulated
% PostScript (EPS) files.  The third figure is a grey scale figure that does
% not exist in EPS form.
%
% Authors have three options for including figure information within a 
% manuscript.  Not all the options may be acceptable by the target Journal - be
% sure to look at the appropriate submission instructions, electronic or 
% otherwise.
%
% Option 1.  Using this option, only the figure captions are included in the
% main body of the manuscript.  The figure captions must start on a new page.
% The captions are generated with the \figcaption[]{} command: the first 
% argument is optional, if you put something in there, put the name of the 
% EPS file that goes with the caption; the second argument is the figure 
% caption itself, and may include a \label command.  The \figcaption command
% generates the figure numbers.  This option is acceptable for all manuscript
% submissions.
\newpage

\begin{figure}[thb]\label{fig1.ps}
{\begin{turn}{0}
\epsscale{0.8}
\centerline{\plotone{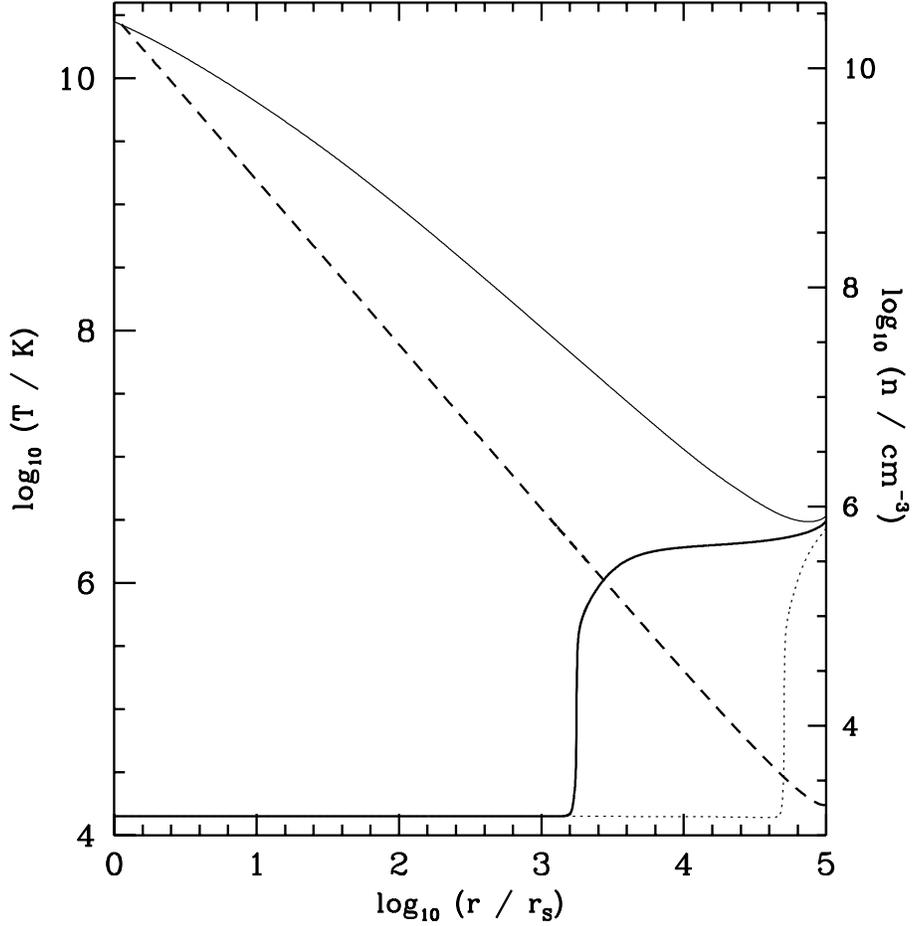}}
\end{turn}}
\caption{Temperature profile for the accreting gas, for three values of the initial
temperature at the outer radius $r_0=10^5\,r_S$ ($\approx 0\farcs08$). 
The accretion rate in all cases
is $\dot{M}=1.5\times 10^{24}$ g s$^{-1}$.  The initial temperature is characterized
by the ratio of thermal to gravitational energy density at $r_0$, which is
$0.19$ (thin, solid curve---the hot branch solution), $0.17$ (thick, solid
curve---the best fit model for M31*) and $0.15$ (dotted curve). Also shown
here is the run of proton density (dashed curve) of the best fit model, whose scale appears
on the right
hand side.}
\end{figure}

\clearpage
\begin{figure}[thb]\label{fig2.ps}
{\begin{turn}{0}
\epsscale{0.8}
\centerline{\plotone{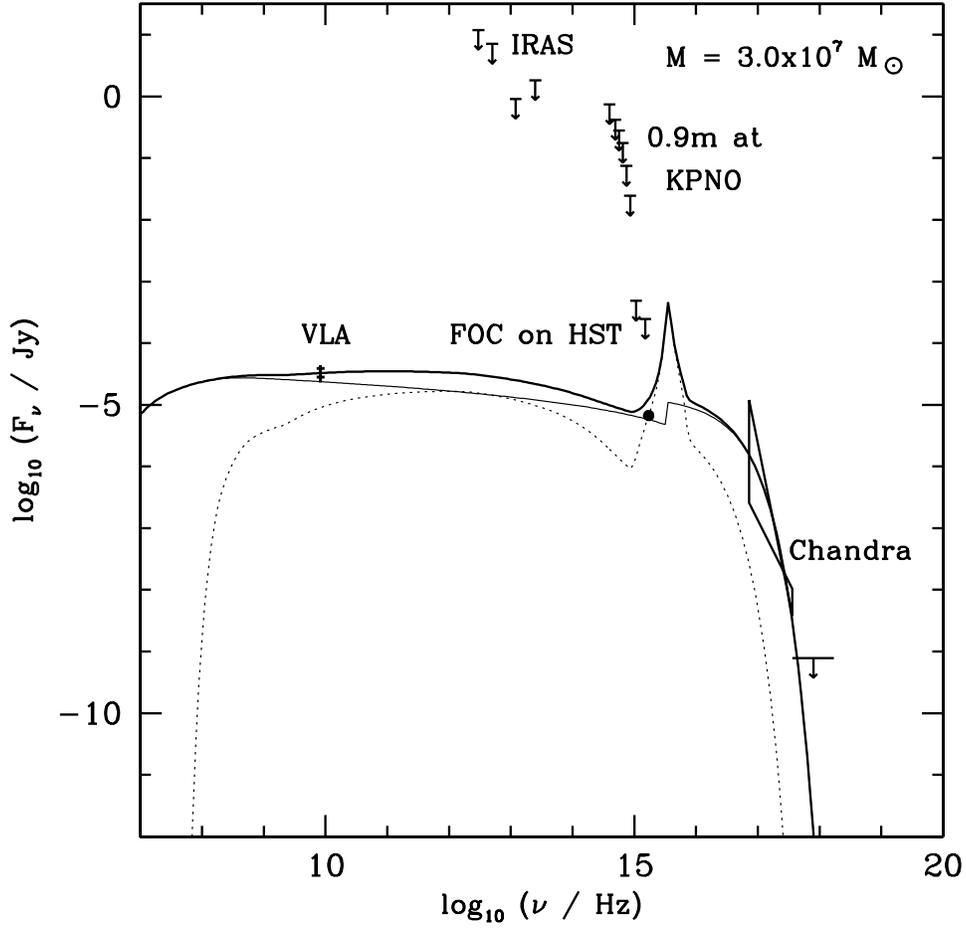}}
\end{turn}}
\caption{Spectrum for the best fit model. The parameter values are specified in
the caption of Figure 1.  Here, the thin solid curve represents the contribution
from the outer hot ($T>10^6$ K) region, beyond $\sim 10^3\,r_S$. The step in the 
curve is due to radiative recombination. The dotted curve corresponds to the
emissivity from the inner cool region, which accounts for the prominent
UV peak due to hydrogen line emission.}
\end{figure}

\end{document}